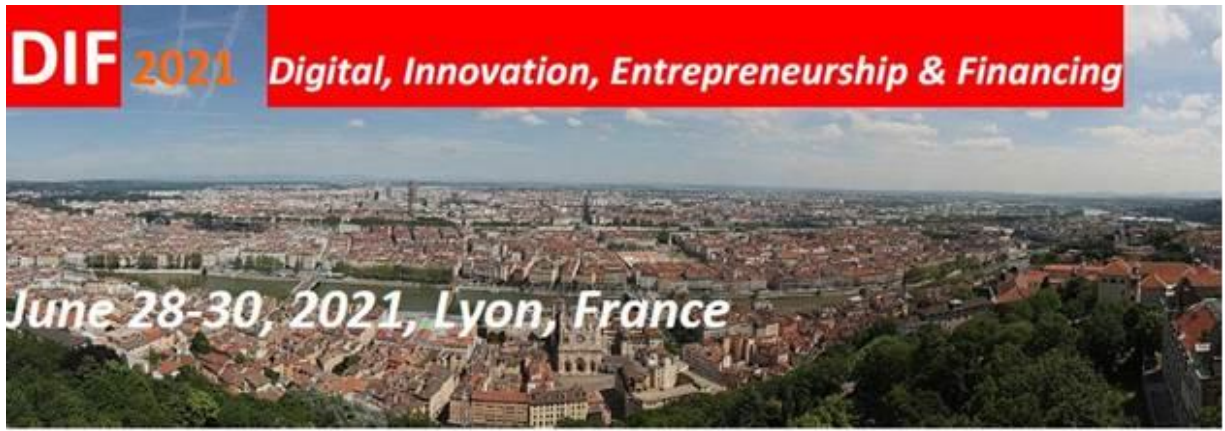

# Tittle: Study of The Relationship Between Public and Private Venture Capitalists in France: A Qualitative Approach.

Jonathan LABBE
*Université de Lorraine - CEREFIGE Laboratory – PhD Student*
jonathan.labbe@univ-lorraine.fr

**Abstract**

This research focuses on the study of relationships between public and private equity investors in France. In this regard, we need to apprehend the formal or informal nature of interactions that can sometimes take place within traditional innovation networks (Djellal & Gallouj, 2018). For this, our article mobilizes a public-private partnerships approach (PPPs) and the resource-based view theory. These perspectives emphasize the complementary role of disciplinary and incentive mechanisms as well as the exchange of specific resources as levers for value creation. Moreover, these orientations crossed with the perspective of a hybrid form of co-investment allow us to build a coherent and explanatory framework of the mixed syndication phenomenon. Our methodology is based on a qualitative approach with an interpretative aim, which includes twenty-seven semi-structured interviews. These data were subjected to a thematic content analysis using Nvivo software. The results suggest that the relationships between public and private Venture capitalists (VCs) of a formal or informal nature, more specifically in a syndication context, at a national or regional level, are representative of an "economico-cognitive" (Farrugia, 2014, page 6) approach to networking and innovation. Moreover, the phenomenon of mixed syndication reveals a context of hybridization of public and private actors that would allow the private VCs to benefit from the distribution of wealth when the company develops its innovation. We can also identify a process related to a quest for legitimacy on the part of the public actor characterized by its controlling role within the public-private partnership (Beuve and Saussier, 2019). Finally, our study has some limitations. One example is the measurement of the effects of relationships on "visible" or "invisible" innovation (Djellal & Gallouj, 2018, page 90).

Keywords: Private equity, public-private partnerships, syndication, innovation, governance.

JEL Classification: G24, O32, O3, G3, M13.

## Introduction.

Financing innovation involves a high level of uncertainty sometimes referred to as "unknowable" (Huang & Pearce, 2015, page 2). In this environment, companies with high growth potential find it difficult to finance themselves. Since the end of the World War II, venture capital then appears for unlisted firms as an alternative to the traditional banking scheme (Andrieu and Casamatta, 2015). Indeed, this model is characterized by its ability to allocate to project holders, technical, technological, managerial, and financial resources in the medium and long term. Following a selection process and in return for the granting of equity, the capital providers will receive "a random gain supplemented by a dividend return" (Wright, 2002, page 283). In turn, the entrepreneur will need to ensure the implementation of cooperative relationships to access new resources (Pfeffer & Salancik, 1978; 2003). Consequently, investors and entrepreneurs constitute the base of relations that initiate the foundations of a transaction materialized by financial and human capital. It is therefore logically towards this perspective that the study of interactions between actors has been oriented (Dubocage & Rivaud-Danset, 2002b; Bonnet, 2005; Guéry-Stévenot, 2006; Barneto & Chambost, 2014; Dubocage, 2016; Panda & al., 2020). However, by observing the development of this mode of financing in France, we can, for several reasons, wonder about the perspective of other relationships: those of public and private venture capitalists (VCs).

The first concerns the specificity of the French private equity model. It was "developed under the impetus of the public authorities and under a logic of public innovation" (Stévenot, 2005, page 27). It constitutes a territorial network where the presence of public and private VCs intersects at a national and/or regional level. In this way, the interactions between these actors and also with other agents can contribute to the creation of innovation networks favoring collaborative exchanges, sharing of information and/or skills (Ferrary, 2006; Djellal & Gallouj, 2018).

Secondly, the use of co-investment operations is another reason. Mainly motivated by risk aversion (Desbrières, 2015), this intervention is qualified as syndication when it is composed of homogeneous actors, only private or public, or mixed syndication, when it includes both public and private investors. In 2019, the proportion of financing through syndication in France is around 60% compared to 37% in the UK (EVCA, 2019). This hybrid arrangement is the subject of particular attention within the existing literature (Murray, 2021). Regarding the observation of the effects of its funding on business innovation. Thus, the studies carried out are mostly based on quantitative analyses (Leleux & Surlemont, 2003; Cumming, 2007; Bertoni & Tykvová, 2012 ; 2015; Brander & al., 2015; Cumming & Li, 2013; Pierrakis & Saridakis, 2017; Cumming & al., 2017; Trabelsi & al., 2019; Awounou-N'Dri & Boufaden, 2020). The main findings obtained indicate that the mixed syndication funding model seems to be the most efficient, especially when the lead is a private VC (Bertoni & Tykvová, 2012 ; Pierrakis & Saridakis, 2017). However, other studies with qualitative or mixed methodologies, conducted in the UK, confirm an efficiency of public VC funding programs (Baldock, 2016; Owen & al., 2019). This research suggests that most quantitative studies in Europe assess the performance of outdated funding models and fail to contextualize the economic development role of these public schemes' (Baldock, 2016, page 1556). In these circumstances, the study of the relationship between public and private VCs finds a legitimate justification.

The observation made earlier characterizes a third cause. "The idea that the public actor is always less effective than the private actor" (Charreaux, 1997; Djellal & Gallouj, 2018, page 87). This observation questions the role of the public actor in relation to the political dimension. To the example of the influence of public VCs in the process of allocating funding within a hybrid organization (Murray, 2021). Or alternatively, the question of the use of public investment programs to support high growth potential firms (Baldock, 2016; Owen & al., 2019).

Finally, there are few studies that focus on investor interactions (Ferrary, 2006, 2010; Ferrary & Granovetter, 2017; Pierrakis & Saridakis, 2019). This is despite the specificities embodied by public and/or private actors in the construction and development of private equity. To date, only one analysis attempts to understand and capture the effects of public-private VCs relationships at a regional level in the UK (Pierrakis & Saridakis, 2019).

Just like the investor-manager relationship we believe that the relationship between public and private VCs is built on formal and informal interactions (Ferrary, 2006; Djellal & Gallouj, 2018). Moreover, this set is marked by a strong informational asymmetry. Consequently, this situation may reveal opportunistic behaviors and/or lead to conflicts between actors (Guéry-Stévenot, 2006; Panda & al., 2020). In the context of syndication, this context can lead to a point of no return with the possibility of the rupture of relations and the eviction of one of the partners. Conversely, we can consider that the relationship between public and private VCs is the site of mechanisms or processes that could promote business innovation. Thus, the objective of our paper can be defined as follows: *How to characterize and define the issues of the relationship between public and private venture capitalists and their effects on firm innovation?*

For this study, we propose to mobilize the public-private partnership (PPP) approach. This perspective will allow us to establish a correlation on the role of the actors in a "strategic" partnership situation. In this way, we will try to identify the incentive and/or "disciplining" mechanisms of the relationship. We also wish to extend this approach by relying on the *Resource-based View* theory. In this way, we will underline the dimension of value conception which is linked to "the notion of competitive advantages and distinctive resources (substantial value)" (Stévenot, 2006, page 185). Indeed, the performance of firms and their ability to innovate will depend on the set of specific resources held by the firm (Pfeffer & Salancik, 1978). Therefore, we hypothesize that exchanges between public and private VCs within complex networks could foster financial and cognitive resources and grant a firm a competitive advantage (Barney, 1991; Ferrary & Granovetter, 2017).

This research is original in that it is one of the first qualitative study carried out on the relations between public and private VCs in France. To do this, we conducted 27 semi-directive interviews with actors in innovation networks identified as being: public and private VCs, research and development directors, company managers and innovation advisors. Our analysis aims to highlight the specific nature of these exchanges in order to propose a definition of the relationship between public and private VCs. We must also take into consideration the occurrence of other phenomena resulting from these interactions. By identifying the crowding out effect as one of them, we will put forward an updated definition of this concept. Furthermore, the analysis of these exchanges may provide elements of reflexion for public authorities. Finally, this research may have some limitations. First, the study of these relationships is part of a complex network perspective where we will try to observe beforehand the effects on "visible" innovation (Appendix 1). This point may be an obstacle in assessing the spillover effects on "invisible" innovation generally present in other types of public-private service innovation networks (Djellal & Gallouj, 2018, page 90). Finally, it seems that to measure the effects of these relationships on business innovation, a quantitative study is otherwise necessary.

Our assisted thematic content analysis indicates that the relationships between public and private VCs of a formal or informal nature, more specifically in a syndication context, at a national or regional level, are representative of an "economico-cognitive" (Farrugia, 2014, page 6) approach to networking and innovation. Moreover, while some VCs retain observable characteristic features in their interventions with companies and managers, the syndication context reveals a hybridization of public and private actors. This would allow the private actor to take advantage of the distribution of wealth when the company develops its innovation. Finally, we can identify a process of quest for legitimacy on the part of the public actor characterized by its control role within the public-private partnership (Beuve & Saussier, 2019).

The plan of our article is organized as follows: in the first part we will present the stages of the relationship between public and private VCs (1). Then we will introduce the context of the relationship by identifying two specific phenomena: the crowding-out effect and syndication financing (2). Finally, we present the choice of a qualitative methodology (3) as well as the results and conclusions of this research.

## 1. The stages of the relationship between public and private investors: from public-private partnership to the theory of complex networks.

In a study conducted in the United States, Ferrary (2006; 2010), states that the relationships between VCs are not governed by any contract. They are informal in nature and take place in a distinct geographical context. Moreover, this environment favors the development of norms that originate in the actions between VCs. To illustrate this fact, Ferrary (2006), proposes to analyze the implicit character of exchanges between actors in a syndication. Using the theory of gift (Mauss, 1930), he shows that the respect of conditions between investors is carried out on pain of exclusion from a network. In a way, these measures contribute to maintaining the balance of relationships. However, they are not intended to solve all the problems encountered by the players. The case of public and private VCs that intervene in an innovation dynamic thus opens new perspectives. In this study, we wish to shed light on the nature of these interactions. By means of the public-private partnership (PPP) approach, we would like to offer the keys to the identification of incentive mechanisms and the definition of the stakes of these investors within the syndication (1.1). One of the objectives will be to verify whether the players benefit from incentives that could benefit the development of business innovation. The framework of this analysis is based on an ecosystem composed of a multitude of heterogeneous actors. This specificity justifies the complex nature of networks (Le Moigne, 1990). Consequently, we will use this approach to determine the attributes of the relationship at the national and regional level (1.2).

### 1.1. Public-private partnership approach: a common definition of the relationship.

The problem of the relationship between public and private VCs can be clarified by the PPP approach in several ways. First, the public-private nature of the relationship within the dynamic framework of innovation can present an interesting orientation. Indeed, this context makes it possible to define a common objective for each of the actors, particularly in the context of syndication financing. If the logics of the different VCs can diverge according to their ambitions, job creation, financial performance and/or conception of innovation, the capacity of the company to innovate remains the main shared representation. Therefore, the issue of innovation can constitute a pattern that orients the interactions between investors towards a partnership logic and leads to the formalization of relationships (Djellal & Gallouj, 2018). In this specific example, we can expect the establishment of a rapprochement from which the actors wish to benefit, benefiting both public and private VCs. This may reflect an initial incentive for the public actor where the lack of competence may symbolize a justification for the use of this type of partnership (Beuve & Saussier, 2019). Especially in areas where private investors benefit from experience and knowledge. Thus, the public VC could compensate for a lack of technical or technological expertise through mixed syndication (Baldock, 2016).

However, this context may present a risk of informational asymmetry. In this example, the VC with the skills can use informational rent to maximize his interests. In this way, the investor can reduce his involvement and adopt a passive approach within the syndication. This situation can therefore lead to opportunistic behaviour, which generally takes two forms: moral hazard or adverse selection risk (Beuve and Saussier, 2019). For all that, control, and monitoring mechanisms as well as incentive mechanisms can "discipline" this conduct. Thus, some contracts may provide for disclosure and exchange of information within a syndication (Admati & Pfleiderer, 1994; Bygrave, 1988). This, all other things being equal, gives a formal perspective to the relationship between VCs. On the other hand, the minority VC may establish control mechanisms of an intentional (Girard, 2001) or spontaneous nature (Charreaux, 1997). Admati & Pfleiderer (1994), show that a contract providing for the maintenance of the same rate of participation in the equity capital during the different rounds of financing can neutralize the impulse of the VC to adopt deviant behaviour. In this way, and thanks to the implementation of these arrangements, the minority shareholder will try to reduce the information deficit that he may experience.

This first illustration also reveals a situation where the private VC would have the capacity to identify incentives relating to the constitution of a co-investment with the public actor. In this case, the incentives can be described as interdependent. The reputation effect, the takeover of a syndication and the question of wealth distribution constitute specific conditions of the relationship between VCs (Wright & Lockett, 2003). For example, the private VC can take advantage of the lack of qualification or experience of the public investor to obtain a form of recognition by the other actors and benefit from a reputation effect. The latter can lead the private equity investor to take a leading position in a syndication. In this context, the reputation of the VC could condition its access to the renewal of future flows and participation in other syndication financing (Desbrières, 2015). Thus, this hybrid relationship could engage in a long-term process allowing the reduction of agency and control costs (Gupta & Sapienza, 1992). The recurrence of financing rounds helps to avoid opportunistic behaviour. Indeed, according to Stévenot-Guéry (2007, page 142), "the dynamic and repetitive nature of the financing rounds" can favor the gain of confidence and reputation within a syndication. Furthermore, "this can encourage parties to share information and seek cooperative solutions when faced with unanticipated contingencies" (Beuve & Saussier, 2019, page 57). Furthermore, if the private VC positions itself as the lead investor in a syndication, it could also be rewarded for its efforts by holding a higher stake compared to minority investors (Lockett & al., 2002). This prospect of future profits may constitute an additional incentive which, in this context, may have a disciplinary effect. This condition introduces the question of the distribution of wealth, which in this example could benefit the private VC.

Another perspective is presented in a study by Leiringer (2006), where the results show that PPPs can acquire stability by bringing together the following four conditions and incentives: freedom of design, collaborative work, risk transfer, and long-term commitment. Under these financial and non-financial conditions, public-private partnerships could provide a favorable environment for the development of innovation. We can establish a connection between these provisions and the motivations for the use of mixed syndication financing. We observe the similar nature of these incentives that may be present within a syndication. Indeed, according to Desbrières (2015, page 3), "three explanatory theories of the co-investment decision by VCs can be mobilized. The first, traditional, is derived from financial theory and sees syndication as a mode of risk sharing via diversification of the portfolio of the VCs (Smith and Smith, 2000). For the second, derived from the *resource-based view*, this practice can be seen as a response by VCs to the need to acquire or share relevant information in the selection and management of their equity investments (Bygrave, 1987; 1988; Brander & al., 2002). Co-investment plays a cognitive role here. From a complementary perspective, which refers as much to the "resources" approach as to the financial approach, co-investment can be seen as a means of increasing their business flows (Lockett & Wright, 2001; Sorenson & Stuart, 2001). In summary, public-private VC relationships in the form of mixed syndication based on a public-private partnership model with specific conditions and incentives would promote and facilitate business innovation.

However, Carbonara & Pellegrino (2020), state that the studies carried out on the characteristics of PPPs and innovation do not make it possible to grasp the attributes that explain this link. To understand and identify the structure of PPPs that enables innovation, the authors identify eight hypotheses (Appendix 2). Using patents as a measure of innovation, they confirm that a structure with a high presence and involvement of the private actor constitutes an organization that favors the development of innovation. One of the reasons that could justify this finding concerns the disposition of the private actor to focus on its production thanks to its technical, technological or financial skills (Beuve & Saussier, 2019). The private VC would have the ability to direct its efforts to ensure that the company obtains an innovation. Otherwise, the public VC, lacking competence could focus on another function such as control (Beuve & Saussier, 2019). Roumboutsos & Saussier (2014), also find that private actors possess an incentive to innovate when the risk managed by them is relatively low. The private VC can in this way, opt for interventions in later rounds to decrease the risk and increase the chances of profitability in return for the investment made.

Secondly, Carbonara & Pellegrino (2020), state that the more partners in the PPP, the less effort they will put in. Under these circumstances, we can again relate this to the passive attitude that can be adopted by a VC. To justify this result, the authors put forward the hypothesis of a non-significant effect of the network dimension on innovation. This finding leads us to question the way in which the networks of innovation financing actors are built or organized. This is to determine the role of public and private VCs in the relations between VCs and also within the interactions with other agents present in these networks. Indeed, we found upstream that networks respect norms that facilitate the balance of relations within a syndication (Ferrary, 2006; 2010). They can also be at the origin of the definition of a common goal following the example of traditional innovation networks (Djellal & Gallouj, 2018). To clarify this aspect, it is appropriate, first, to distinguish between the forms of relationships between VCs. In this way, we will try to identify whether a geographical or economic specificity is linked to the organization and constitution of this ecosystem. To achieve this, our approach will be based on the theory of complex networks.

## 1.2. Complex network approach: keys to a distinct national and regional relationship.

The complex network approach (Newman, 2003; Newman & al., 2006) aims to understand and apprehend the role of the multiple actors involved in an ecosystem. Thus, in our study concerning the relations between public and private VCs, we do not wish to reduce the analysis of the various interactions to these protagonists alone. We hypothesize that to explain or describe a phenomenon, more particularly in the context of innovation, it seems essential to consider "the intervention of several agents characterized by the non-linearity of their interactions" (Ferrary, 2010; Ferrary & Granovetter, 2017, page 327). These different actors can contribute to the constitution of innovation networks and allow through collaborative exchanges, the sharing of information or skills (Ferrary, 2006; Djellal & Gallouj, 2018). For all that, it is conceivable that disparities appear in the formation of these networks. And this, due to distinct geographical and economic contexts. Thus, we will attempt to present the characteristics of the relations between VCs from a regional and national point of view in France, then we will specify the "market" or "non-market" interest of these exchanges (Djellal & Gallouj, 2018, page 88). Thereafter, we will propose a definition of the relationship between public and private VCs.

In France, private equity financing of companies has mainly "developed under the impetus of the public authorities and under a logic of public innovation" (Stévenot, 2005, page 27). Several reforms have been carried out and have contributed to the construction of a territorial network with both a public and private presence. One of them, in 1985, was a major step forward with the creation of SCRs (venture capital companies). In this way, the State "institutionalized venture capital as a means of financing business development" (Stephany, 2001, page 65). Consequently, the policy of supporting innovation has become an opportunity to encourage a change in attitudes towards entrepreneurship.

To consolidate these objectives, since 2013, a complex territorial network has been structured around major players such as the public investment bank[1] (BPI) or the Caisse des Dépôts[2] Group (CDC). For these entities, one of the first actions was to focus on the professionalization of certain players and activities within private equity. At the same time, these institutions aim to attract private investors and increase fundraising for innovative companies. This restructuring of economic activity is a specific stage at the national level and within different regions for which public and semi-public private equity firms are key players (Leleux & Surlemont, 2003).

However, this attempt to balance the presence of public and private actors on the territory may present inequalities. These inequalities are mainly due to the historical, economic, and institutional development of each region. For example, the historical origin of the increase in private VCs in these areas is often linked to the families of industrialists who, for the most part, formed large national groups. Thus, some family firms participated in the setting up of private regional structures or in the creation of SCRs that still exist. At the same time, other players such as banks have also contributed to the development of this activity in France. These different institutions, present locally and nationally, will represent a model in terms of financial performance as well as for other specific skills (due diligence, etc.).

In a study conducted in the UK, Pierrakis & Saridakis (2019), will using an online survey, highlight the inequalities of public and private presence in some regions. The authors reveal that public venture capitalists have "a greater role to play in mobilizing the different actors in the regional innovation ecosystem" (Pierrakis & Saridakis, 2019, page 850). A first explanation is that early-stage business development is more supported by public VCs in certain spaces (Mason & Pierrakis, 2013). It requires more interactions due to specific financial and non-financial resource needs. This finding is also made by Mason & Pierrakis (2013) and Murray (2021), who point to an increasing increase in UK public venture capital intervention in the seed stages since the burst of the internet bubble. While this inequality may lead to complications, for Sorenson & Stuart (2001), the issue of these regional interactions may be an important step in creating new investment opportunities. This is because these exchanges can affect the spatial distribution of financing activity as social relations tend to cluster in geographical and social spaces (Sorenson & Stuart, 2001; Pierrakis & Saridakis, 2019). Thus, Pierrakis & Saridakis (2019, page 855), will define these geographical spaces as "networks that comprise a set of horizontal and vertical relationships and with other strategically important actors" (Gulati & al., 2000) (Appendix 3).

---

[1] The BPI, a public investment bank, was created on [1] January 2013. It brings together the entities, FSI regions, Oséo and CDC. Its objective is to support SMEs, ETIs and innovative companies.

[2] Caisse des Dépôts et Consignations was created on 3 July 1816. Originally, its mission was focused on four main areas: consignments, voluntary deposits, retirement funds and complementary activities. It has been contributing to the objective of supporting growth and territories since the 1950s. In 2013, it included the public investment bank among its five business lines and recently, in 2018, the bank of territories. It represents a network of 35 operational centers that mainly support innovation financing in the regions.

Based on this hypothesis and the specificities of the French venture capital model presented above, we emphasize that the relationships between venture capitalists are not limited to these two actors. They include the existence of other protagonists (company managers, research and development directors, engineers, etc.) and heterogeneous structures (research and development centers, incubators, business development organizations, etc.). The latter also participate and contribute to the support of innovation. Thus, for Djellal & Gallouj (2018), innovation networks should not be limited to the sole presence of market activities. We must also take into consideration the non-market activities involved. Not taking these into account can constitute a bias "at the origin of a triple *gap*: an innovation *gap*, a performance *gap* and a public policy *gap*" (Djellal & Gallouj, 2010; 2018 page 81). As a result, a new contrast both geographical and economic can be observed, this time due to the inequality in the clustering of VCs, agents, firms and market or non-market activities.

From this observation, we can suppose that certain French regions can let appear networks where the multiplicity, the complementarity and the interdependence between agents can give rise to innovative clusters which associate researchers, entrepreneurs and investors (Saxenian, 1994). The complex network approach seems to indicate that relations between VCs do not simply take place on two distinct levels, national and/or regional. They are carried out in territories presenting an entrepreneurial, financial, and social concentration. Therefore, the presence of VCs in clusters would be likely to foster interactions with other agents, which represents "a characteristic dynamic of innovation" (Ferrary & Granovetter, 2017, page 328). Porter (2000), defines the cluster as a combination of competitive and cooperative relationships between co-located firms allowing for better learning. If the structuring of these networks allows public and private VCs to establish the same type of relationships identified earlier, this context can encourage the promotion of the concept of innovation through cooperation and collective learning within them (Leducq & Lusso, 2011). However, the cluster should not be seen as too locally rooted. This is the observation made by Bresnahan & al, (2001, page 835), who indicate that "the economic factors that initiate a cluster may be very different from those that sustain it". Adams & al, (2018, page 879), also state "the same is true for the institutional support associated with economic factors when the center of a cluster shifts over time". Thus, an innovative cluster is not intended to be territorially anchored and may undergo geographic, economic, and institutional evolutions or transitions. Consequently, the relationships between public and private VCs within an innovation network will evolve according to these same factors. These circumstances can lead to inequalities in a territory. Nevertheless, it seems that the fact that public and private VCs belong to a complex network could favor the effects of their relations on the innovation of enterprises.

To understand the establishment of relationships between VCs within networks, we can try to define their role(s) and characterize some specific functions. According to Ferrary (2010); Ferrary & Granovetter (2017), interactions with other agents allow SCRs to support innovative projects and lead to a value creation process. To this end, they identify five main functions performed by VCs within complex networks: funding, selection, signaling, collective learning and rooting. To illustrate this, we can take the example of the process of evaluating a company. This step will constitute an essential and repeated dimension in the foundation of formal and informal interactions between VCs, between public and private VCs and with other agents. It will occur upstream and/or during all investment rounds. Through this process, a VC can contact other investors to obtain information on a project, a company, an entrepreneur, and/or obtain technical or technological expertise. This can promote information sharing and knowledge transfer between agents (Kogut & Zander, 1992; Burkhardt, 2018). Consequently, the creation of new competencies occurs through this learning process and through cooperation between actors. Moreover, this stage offers the possibility of forming a partnership or strategic alliance between private equity firms (Ferrary, 2010; Burkhardt, 2018). Thus, this phase can bring about elements of convergence that may lead to co-investment. For example, during the evaluation process, an investor with industrial expertise recognized by the other players can convince them to carry out syndication financing. In this way, this method can be likened to a condition for sending a signal to other investors (Akerlof, 1978). Once the company is financed, individuals can conform to an organizational principle that materializes in the form of internal collaborations (Kogut & Zander, 1992).

The support phase also allows for a more in-depth study of the interactions between investors and between investors/managers. This stage provides details on the conditions of cooperation between actors. It is also in this phase that the strategic dimensions, trust, power, embeddedness have an importance within the relations between VC and VC/manager. For Guéry-Stévenot(2006, page 159), it is "in the interactions between the shareholder and the manager that decisions are formed and the construction of knowledge and the evolution of mental schemes are played out". Thus, the quest for recognition, status or power is indicative of dyadic relationships between actors and the embeddedness of these networks (Granovetter, 1985; Leducq & Lusso, 2011). For Bygrave (1988, page 155), "collaborations or exchanges between actors characterize the position of individuals in social networks that can facilitate or impede economic action" (Granovetter, 1985). Consequently, the individual dimension in these networks can therefore exert constraints on the innovation process (Granovetter, 1985).

In the context of this study, the complex network theory is useful for us to identify that "the dynamics of an innovation network is founded by a multitude of agents of which public and private VCs constitute a robust foundation thanks to their capacity for anticipations and learning that support this system" (Ferrary & Granovetter, 2017, page 339). The characteristic functions of public and private VCs allow them to establish formal and informal relationships, between VCs and other agents that could contribute to the value creation process. Indeed, the mobilization of networks to collect information can define a perimeter for knowledge exchange and resource sharing (Penrose, 1959; Ferrary, 2006). To conclude this part, we can propose the following definition of the relationship between public and private VCs:

> A set of formal and/or informal relationships between heterogeneous investor-capitalists, heterogeneous investor-capitalists and agents that facilitates the exchange of cognitive and financial resources and is part of a value creation process.

We will now address the context of relationships between VCs by seeking to identify the reasons for the occurrence of a specific phenomenon: the *crowding-out* effect (2.1). By combining an approach based on the hybrid character of syndication, we will propose an update of the definition of the crowding-out effect. We will then mobilize the *Resource-based View* approach to determine the conditions of value creation levers within a mixed syndication (2.2). Thus, our objective will be to apprehend the mechanisms or processes of value creation and wealth distribution through these interactions.

## 2. From the crowding out effect to the syndication relationship: a perspective from the hybrid nature of syndication to the resource-based view.

The financing of innovative enterprises is linked to a strong informational asymmetry. In this context, relations between public and private VCs may be subject to conflict. In the context of co-investment, these disagreements can be the cause of a relationship breakdown: the crowding-out effect (2.1). Other reasons, such as institutional dissimilarities arising from the characteristics of public and/or private actors, or the strategic nature of certain financing, may be at the origin of this effect. In this section, we will also try to verify that the various resources and capabilities brought within a mixed syndication (2.2), which is here considered as a hybrid form of organization (Murray, 2021), allow for value creation (Borys & Jemison, 1989; Villani & al., 2017).

### 2.1. Definition and presentation of the crowding-out effect: cause and consequence of the relationship between public and private actors.

Using a macroeconomic perspective with the effect of public spending as the original factor (Spencer & Yohe, 1970; Aschauer, 1989), the existing literature defines the *crowding-out effect* as the absence of private investors characterized by public intervention (Spencer & Yohe, 1970; Aschauer, 1989; Cumming & MacIntosh, 2006). Complementing this research, recent analyses related to private equity financing of companies suggest that this process can occur at two levels: the market and/or the firm (Leleux & Surlemont, 2003; Cumming & MacIntosh, 2006; Brander & al.,2010; 2015; Cumming & al., 2017). For example, Brander & al. (2015, page 573), state that "the crowding-out effect at the firm level means that the publicly funded firm receives less private funding". However, while the terms of crowding out are at the firm level, these conditions do not necessarily hold at the market level. Whether from a microeconomic or macroeconomic point of view, the literature therefore considers this phenomenon to be fully linked to financial capital and resulting from the action of the public actor. Based on this observation, we can raise the possibility of developing a contrary hypothesis: an investment made by a private VC can crowd out, in whole or in part, a public VC. In this respect, during a co-investment, situations of disagreement may decide a private or public VC to pursue a financial intervention independently. Thus, the hybrid nature of the mixed syndication phenomenon does not constitute a firewall for the *crowding-out effect*. These conditions can be seen as a first update of the definition of the crowding-out effect.

Furthermore, we must also consider the concept of human capital which, as we have seen previously, can be a unifying element in the relationship between VCs. We can imagine that in contrary conditions, this concept could intervene as an exclusion factor. Indeed, if relationships can within a mixed syndication promote and facilitate the exchange of skills or knowledge (Mahoney & al., 2009; Murray, 2021), the opposite effect is also possible. For example, interactions between investors can lead to conflicts regarding a lack of technical, technological, managerial, or financial skills. It can also be noted that in the absence of skills and in order to avoid exclusion, this situation can lead some actors to a phenomenon of mimetic or strategic isomorphism within a syndication (Dubocage, 2006; Dubocage & Galindo, 2008). This finding implies that the public and/or private VC would be able to lose its reputation within an innovation network. We can therefore see that the crowding-out effect can act as a mechanism for *ex ante* and *ex post* disruption of the relationship between public and private investors.

If mixed syndication constitutes a hybrid form of organization it can, however, combine different institutional logics. This representation, which can take the form of a public-private partnership, can be at the origin of forms of collaboration that enable value creation (Villani & al., 2017). However, the distinct institutional logics of public and/or private actors, can be the source of tensions or sources of conflict (Bishop & Waring, 2016). These disparities can be seen in the direction of strategic financing choices made in a company. According to these modalities, these institutional differences within a hybrid organization could be the source of a crowding out effect and lead to the breakdown of relationships between public and private equity investors. However, the context of repeated negotiations that the mixed syndication framework would offer would be favorable to the resolution of conflicts related to the institutional logics present in hybrid organizations arranged as a public-private partnership (Bishop & Waring, 2016). According to Villani & al. (2017), the conditions enabling value creation are determined by the ability of public and private actors to make strategic choices taking into account organizational and institutional specificities within a public-private partnership. We can hypothesize that mixed syndication would constitute this hybrid organization. Consequently, it would bring together "the existence of divergent representations which could favor cognitive progress and the discovery of innovative and relevant solutions" (Stévenot-Guéry 2007, page 171) as well as strategic decision-making. Considering all the elements presented in this section, we can propose the following definition of the crowding-out effect:

> The *crowding-out effect is* a temporal effect resulting from the intervention of a public or private actor in the form of financial or human capital and leading to a partial or total crowding-out of public or private actors at company and/or market level.

In this second part, we will try to demonstrate that financial and cognitive resources within the relations between public and private VCs can be a determinant of value creation. For this purpose, we consider that the firm is here constrained and dependent on public and private VCs and other agents of the complex innovation network that control these resources (Pfeffer & Salancik, 1978; 2003).

## 2.2. The syndication phenomenon: a justification for the mobilization of cognitive and financial resources.

The complexity of the representations of public and private actors within syndication brings together resources that are difficult to obtain, sell, imitate, or replace placing strategic value on them (Amit & Schoemaker, 2016). As a result, this form of public-private partnership can provide valuable, scarce and non-substitutable skills and resources (Barney, 1991). Complementing this approach, we can consider the assumption that the performance of the firm, here represented by innovation, depends on the specific resources and skills generated within the mixed syndication. Thus, "innovation would be encouraged to the extent that strategic control is in the hands of those who have the incentives and capabilities to allocate resources to uncertain and irreversible investments" (Miozzo & Dewick, 2002; Roumboutsos & Saussier, 2014, page 349).

Based on this observation, the enterprise can be perceived as a place where individual knowledge is transmitted by means of learning phenomena and can generate a competitive advantage (Spender & Grant, 1996. In the same way as cognitive resources, the financial resources provided because of mixed syndication financing can improve the firm's chances of succeeding with its innovation project. In this way, the concentration of the resources obtained can "guarantee" the creation of value and ensure the identification of opportunities and the neutralization of threats (Barney, 1991). In the same way, a long-term relationship would allow the company to accumulate resources from management skills, themselves derived from the experience of public and private VCs (Penrose, 1959). It is this accumulation of knowledge that gives a dynamic process to the relationship between public and private investor capital. For example, Quas & al,(2020, page 1), note that "the most experienced venture capital firms bring valuable non-financial resources". Finally, the heterogeneous nature of the resources will enable the firm to gain a sustainable competitive advantage (Penrose, 1959; Peteraf, 1993). This condition can be achieved by the fact that public and private actors will be able to contribute varied knowledge, skills, or financial resources. However, we must also take into account the coordination costs that may be incurred by the firm due to different cognitive patterns (Langlois, 1992). Nevertheless, learning processes or skill transfers can facilitate the alignment of actors which can prove to be an advantage for the firm.

## 3. Methodology.

To understand the study of the relationship between public and private investor capital, we first sought to clarify our research question (Trudel & al., 2006). We then considered the empirical context of our research. This seems to direct us towards a methodology that "assumes that we see actors talking, acting, interacting, cooperating and confronting each other" (Dumez, 2011, page 49). In this way, we can attempt to characterize the representations of the different VCs within the relationships. We also note that a quantitative study alone, would not capture the representations of the venture capitalists, especially during their interactions. Moreover, research relying on a quantitative methodology is intended to focus mainly on variables and not on actors (Dumez, 2016). Consequently, we have chosen a qualitative methodology with an exploratory aim. Thus, we will now introduce the context of our analysis (3.1) and present the choices made for the collection and processing of our data (3.2).

### 3.1. The choice of a qualitative study: context and presentation of the study.

Our research approach is, first and foremost, linked to the desire to establish meetings with venture capitalists to initiate work on the relations between public and private investors. Therefore, we first participated in meetings of associations belonging to innovation financing networks. In this way, we observed the participants "thinking, speaking and interacting in a situation" (Popper, 1979). We also needed to obtain contacts who could recommend us to other potential actors. To do this, we repeated this process within several networks, present in different regions, to obtain a variety of actors within our sample.

Thanks to this mechanism, we set up three initial interviews with regional VCs. The objective of these meetings was to develop our reading grid in terms of refining our problem and the questions that could be mobilized. We also took the opportunity to interview simultaneously public and private VCs that had collaborated on projects and/or had carried out syndication financing. In this way, we wished to highlight our "qualitative approach, which only makes sense if it shows and analyses the intentions, discourses and actions and interactions of the actors, from their point of view and the researcher's point of view" (Dumez, 2016, page 15). Consequently, this introductory phase can be characterized as an exploratory process. Moreover, the existing literature and the first interviews conducted confirmed that in order to study the relationships between public and private VCs, we also needed to collect several interviews with actors from complex innovation networks (Ferrary & Granovetter, 2017; Pierrakis & Saridakis, 2019). In order to identify the protagonists present within these ecosystems, we chose stakeholders represented in the diagrams of Ferrary and Granovetter (2017), (Appendix 4) and Pierrakis & Saridakis (2019), (Appendix 3). Thus, we have selected: private, public, semi-public VCs, company managers, research and development (R&D) center directors, innovation consultants as well as venture capital lawyers (Ferrary, 2010; Ferrary & Granovetter, 2017; Pierrakis & Saridakis, 2019). We will detail below, the criteria behind this selection.

To characterize the relationships between public and private VCs, we first had to obtain a list of these players. To these forms of SCR, we chose to associate semi-public VCs, which are essential players in the private equity business in France, just like public or private ICs (Leleux & Surlemont, 2003). In addition, there is a second objective, which consists in perceiving the effects of these relationships on the innovation of companies. For this, we retain here, the definition of "visible" innovation presented by Djellal & Gallouj (2018) (Appendix 1). Indeed, the private equity industry has a specific interest in technological inventions made visible through different variables or measures of innovation (Hagedoorn & Cloodt, 2003). In this respect, our choice was oriented towards research and development (R&D) expenditures and patent filings, which are data mainly mobilized within the existing literature on innovation financing (Griliches & al., 1991; Bertoni & Tykvová, 2015; Pierrakis & Saridakis, 2017; Awounou-N'Dri & Boufaden, 2020; Cumming & al., 2020). Under these conditions, we also selected other actors related to this environment such as R&D managers and technological innovation consultants. Moreover, the collection of interviews with managers of technology companies belonging to fintechs or biotechnologies results from this same logic (Dubocage & Galindo, 2008; Gazel & Schwienbacher, 2018; Awounou-N'Dri & Boufaden, 2020). Finally, the legal context of these relationships, especially in the context of mixed syndication, can provide us with essential elements regarding the definition of contracts, incentives, or specific conditions observable during exchanges between public and private actors (Cumming & al., 2010; Pierrakis & Saridakis, 2019; Cumming & al., 2021) (Appendix 3). Therefore, we decided to conduct interviews with venture capital lawyers.

In our sample, we will analyse the cases of companies in the startup or development phase. Thus, the situation of these companies can allow us to observe the way in which investors proceed to mobilize their network(s) upstream, during the process of evaluating a project as well as during the support phases. We can also obtain an overview of the function of each of the investors in an *ex-ante* and *ex post* financing situation. The effects of mimetic isomorphisms or collaborative learning (Dubocage, 2006; Guéry-Stévenot, 2006; Dubocage & Galindo, 2008; Granz, 2021) may also appear in these early rounds where interactions between investors are likely to be more numerous than at other stages of financing. This stage leads us to the investment phase where the VC can choose to finance the company alone or in syndication. To these criteria, we add their geographical location: regional or national investors (Ott & Rondé, 2019; Pierrakis & Saridakis, 2019; Alperovych & al., 2020; Van Aswegen & Retief, 2020; Boyer & al., 2021). As well as the size, or type of fund selected. We also sought to obtain interviews with national actors or from other regions. For this, we mainly mobilized recommendations as well as social networks (*LinkedIn, slack*). To characterize this last aspect, it seemed essential to us to conduct interviews with public and private VCs that are or have been involved in co-investments or syndication financing. Consequently, these interviews can reveal the interplay of actors and specify the conditions of agreements or disagreements. Moreover, these interactions, constitute "a choice data because they represent a link, even temporary, between the actors of a network which, repeated or not, can otherwise promote relationships of influence, power or collaboration between them" (Adrot, 2019, page 285). In this way, "interaction constitutes material that allows us to analyze networks and reveal the collective dynamics or structural characteristics of certain collectives" (Scott, 2017; Adrot, 2019, page 285).

To respond to our problem, semi-directive interviews were conducted with a number of stakeholders. By mobilizing this technique, we wish to create an exchange and obtain spontaneous developments from the various protagonists interviewed. The objective is also to collect the perceptions of the actors in a given context, in order to gather rich and varied data offering explanatory power for the processes (Miles and Huberman, 2005). For all that, we also ensured that we respected the theoretical saturation criteria that are obtained when the data collected no longer provide new information (Wacheux, 1996; Saunders & al., 2018). In total, we conducted 27 one-hour interviews with national and regional actors from 8 region: Auvergne-Rhône-Alpes, Bretagne, Grand-Est, Hauts-de-France, Île-de-France, Nouvelle Aquitaine, Pays de la Loire, Provence-Alpes-Côte d'Azur. We chose to meet the interviewees at their place of work. Thus, sixteen out of 27 interviews were carried out on this model. Of the remaining 11 interviews, 8 were conducted by video conference and 3 by telephone. We would like to point out that two interviews were conducted with in-house counsel who wished to withdraw and no longer share the data that had been collected. We were unable to obtain further interviews with this category of actors.

### 3.2. An assisted thematic content analysis.

For all the semi-structured interviews conducted, we made a complete transcript. These data enabled us to carry out a thematic content analysis. Indeed, our approach is interpretative and proposes an analysis of the actors' discourse whose repetition of expressions, metaphors or identification of certain perceptions can help us characterize certain specific mechanisms. This posture can facilitate the understanding of the meaning brought by the actors to describe the context in which they find themselves. Moreover, by mobilizing this approach we wish to limit as much as possible certain biases, such as the confirmation bias which consists in sorting the interview data to obtain validation of certain hypotheses (Budin & Romelaer, 2019). Finally, the thematic content analysis assisted by the NVIVO software allowed us to process the "material independently of the theoretical framework of departure" by means of coding (Dumez, 2016, page 57). In this way, we proceeded to segment the content of our interviews into units of analysis that we integrated "within categories selected according to the research object" (Averseng, 2011, page 377). We will now, introduce the results of our qualitative study.

# 4. Results of the qualitative study.

The analysis of the interviews seems to define the relationship between public and private VCs as an "economico-cognitive" approach to networking and value-creating innovation (4.1). In addition to this characterization, this research highlights the particularities of different types of investors. Moreover, this last element brings clarifications on the question of the distribution of the created value. Our study also evokes the context of mixed syndication, which may be representative of a hybrid form of organization favoring the exchange of resources and skills (Murray, 2021) (4.2). Finally, within this analysis, we note that the innovation network dynamics in France may be marked by a "Silicon Valley" effect (Ferrary & Granovetter, 2009, 2017). Under these conditions, the actions of the public VC may specify a quest for legitimacy through control with other agents and underlines the importance of the political dimension in its interventions (Murray, 2021) (4.3).

## 4.1. An "economic-cognitive" approach to networking and innovation that creates value within relationships and the company

Relationships between public and private VCs of a formal or informal nature, in a syndication context, at a national or regional level, are representative of an "economico-cognitive" approach to networking and innovation. According to Farrugia (2014, this approach embodies a strategy of knowledge capitalization that confers economic surplus value to the individual who holds it. It is not limited to this first conception and can also represent a form of accumulation of financial resources. The issue of this capitalization is "symptomatic of embodied social dispositions to accumulate and make individual or collective entities bear fruit" (Farrugia, 2014, page 6). Consequently, if the ambition is to make the financed company grow and prosper, this approach may also concern the co-investment actors and more particularly, the syndication leader. In this context, the incentives that justify the involvement of VCs in this position may differ. For the private VC, the use of mixed syndication would consolidate the guarantee of financial performance. Conversely, the public VC would focus on the economic dimension related to the social benefits that can be derived from the profitability of the enterprise. To achieve these different ambitions, this approach is based on the concept of innovation. It plays a central role that can be identified here as an incentive mechanism to share specific resources in a logic of value creation. The foundations of the relationship can thus be based to a large extent on this aspect. **CIPri 5** describes this approach as follows:

*"Innovation is the determining criterion. Our friends from the CDC with whom we invest, they put funds with us only in this logic. In any case, this is the prevailing logic today. The objective is to focus above all on innovation with our systems and our businesses. In our region, we are really looking for support for start-ups, and so the first part of being able to make a co-investment is to be in phase with an innovation reference system that we have determined, and which allows us to determine a whole bunch of criteria to say "ok, this is an innovative company, we can call the CDC".*

Beyond the concept of hoarding, this singular approach is based on another essential characteristic. This is the network dimension. It acquires a particular meaning, in the context of co-investment or mixed syndication financing (Pierrakis and Saridakis, 2019). Even if this phenomenon is observable for national and/or regional VCs, it is accentuated most often in the regions thanks to the proximity of investors and heterogeneous actors present in what can be described as an ecosystem or innovation park (Pierrakis & Saridakis, 2019; Boyer & al., 2021). Moreover, public-private syndication seems to send a signal to other investors or actors in these dynamic networks in the context of a possible collaboration. **CIPub7** thus finds that:

*"The big advantage of doing only public-private co-investment is the network. In our region, most of our activity is based on it. When I say network of investors, it's a bit limiting, we work in a network. And we have the great advantage of opening doors thanks to the public-private partnership. We can work in a rather ecumenical way with a certain number of other investment funds with which we try to be on good terms. I'm also talking about the banking network, the accountancy network, all the people that there may be... the consular network and all the people who may be in contact with a company at a given moment and have a good perception and knowledge of it, we really try to make this regional network live".*

In general, a VC can establish formal or informal interactions to obtain information about a project, an entrepreneur, another investor (Ferrary, 2006, 2010; Ferrary & Granovetter, 2017). For this, he will mobilize his personal and/or professional network which can be constituted, of experts, managers, academics, investors. This dimension on which the public or private investor capital relies has a double function: obtaining and circulating quality information. According to **CIPri 2**, it is common practice to carry out what is identified in the work of Ferrary (2006) as a gift for a gift:

*"We rely on the network to benefit from information from a maximum number of investors or experts. If there is a complex case or certain files that are too technical, we can call on professionals who can give us an expert opinion on the technology, we rely on the external network just as other investors can obtain information from us in exchange for a service.*

*As far as* the capitalization of knowledge is concerned, the context of mixed syndication formalizes the relations between VCs and seems to confirm "*a willingness to exchange on each other's know-how in order to solidify what can be qualified as a shared experience*" **CIPub2.**

The data presented in Table 1 indicate that this "economico-cognitive" process constitutes a singular approach with a collective effect which takes the form of a financial collection, and which encourages the capitalization of knowledge. This approach is carried out under a logic of common sense represented by innovation and whose main objective is to "make the innovation of companies bear fruit". However, private investors under this logic covet a future remuneration. This aspect reveals a strategic determinism where the capitalization of knowledge and financial resources seems to constitute an incentive for the private VC to take the lead in syndication. This situation clarifies the question of the distribution of wealth that can, under these conditions, benefit the private investor.

*"It is true that when you have money and experts around the table, it is in your interest to have more collaborative relationships. As a result, we have a strong interest in directing co-investment. With the idea that we must not lose sight of the fact that it is a strategic partnership. For us, it represents a financial partnership that must be respected and if the company can achieve growth as planned, our network will be grateful, and the performance will be there"* **CIPri 3.**

| Empirical Results | Verbatim |
|---|---|
| **An "economic-cognitive" approach to the relationship: approaches, actors and more?** | |
| Mixed syndication a singular approach with a collective effect... | "We rarely invest on our own because our relationships with the private investment community mean that we systematically engage with them. We want to call on them so that we don't do the files alone, we want a collective effect on what we finance" **CIPub 1.** |
| in the form of a financial collection... | "Let's take the example of a company we funded with **CIPub 7** in the first round. We raised the funds in our network by engaging in co-investment. The manager needs a second round and I expected this, I told the manager to get his balance sheet out and prepare a business plan quickly. We looked at it last week and this afternoon, I'm taking him to **CIPub 7** and there I'm taking him directly to determine the sum to be raised so that we can work together on the second round, that's how we work in the region between public and private players " **CIPri 4.** |
| and in the form of a capitalization of knowledge... | "By making co-investments, we have a combination of skills and information, we can't say that we have a scientific approach but rather a financial one, but as time goes by we all start to know a little more together and that inevitably when we accumulate, it benefits the company that we have financed" **CIPri 1.** |
| with a commonsense logic and a specific strategy... and shared... | "We also have, I think, this kind of state of mind of feeding and confronting experiences and ideas, these places where we can share and exchange. All this at a given moment offers us a whole series of regroupings to invest together and to ally ourselves to encourage the development of innovation ". **CIPub 4.** |
| An innovative approach with a future remuneration objective. | "Innovation is everything that a company does differently from what it did before and that will allow it to earn more money. So, innovation is technological, it's managerial, it's financial, it's whatever you want ... It's everything we do differently from before to earn more. I like this definition and it's our way of working" **CIPri 2.** |

We have seen above that private equity can develop a strategic determinism where knowledge capitalization can be an incentive to take the lead in syndication. However, the "economico-cognitive" approach also implies a logic of common sense and the expectation by other investors of an alignment of interests and values. In order to maintain its reputation and legitimacy within syndication, the investor-capitalist will sometimes have to give up this power grab (Charreire-Petit & Dubocage, 2018).

*"Shared values is a nice term, it's better when we are around the table sharing good practices. I think that in certain situations we are better off when we don't try to control everything and instead try to obtain a collaborative partnership that may include some sacrifices"* **CIPri 6.**

This first part seems to show that the relationship between public and private investor-capital can be defined and characterized as an "economico-cognitive" approach to networking and innovation. The mixed syndication constitutes the place of a shared representation symbolized by innovation where the capitalization of financial and cognitive resources can be value-creating. Following these first results, we wish to understand the representations of public and private VCs outside and in the context of mixed syndication.

## 4.2. Characteristic features and hybrid nature of syndication.

Within the relationships between public and private VCs, perceptions may arise that are linked to the specificities of the actors. These may result from behaviour, practices, or from the reputation endorsed by other actors. However, our thematic content analysis shows that all the actors in the innovation network characterize public and private actors according to distinct logics. This characteristic can lead to a problem of alignment of cognitive schemas (Langlois, 1992). For private VCs, the representation of the public VC is above all linked to institutional symbolism. The public VC is perceived as an agent who works for the "*government, which sets both the objectives of these investors and their incentives through a remuneration that is not subject to the same vagaries as private VCs*" CIPri2. CiPri 2 also adds that "*the personal motivation of public VCs is linked to the political dimension and the reputational stakes of the institution. We can take the example of files that we do not want to finance, the public manager does not want to finance either, but the State will decide, because it is a particularly sensitive file, so the State wants them to invest and well... these are characteristics that make it a little different and a little less attractive for us private investors*.

This institutional and political dimension is a specific characteristic of the action or intervention of public VCs. Moreover, even if the actors in the innovation network who interact with public and private VCs grant certain representations or institutional features to the latter, it seems that the dimension of trust and, more particularly, interpersonal trust limits their effects.

*"That is, we know the role of the private equity firm, especially if it works for a bank. We know the institution and its credentials. But I think that our interlocutors do too. We know very well who we are dealing with and if this is the case, we will work with this partner in a very collaborative way at some point it must correspond to our mutual interests, with an*

*abstraction of our differences and a pooling of our complementarities, for this we trust the person in front of us "* **ConseillerInnov2**.

However, the implementation of public policies in support of innovation funding and the financial needs of enterprises seems to confer the role of public administrator on government VCs. This representation appears to be an issue for some managers and a necessary process for validating a step that will subsequently be recognized by other players.

*"We created our company in 2007, and in 2008 when we knocked on several doors of private investors, we were told that it was not the time for our innovation. We quickly got in touch with the public structures, we knew that they would help us especially during the crisis, and it was the policy at the time to help the creation of companies and even more so if there was an innovative project. We knew that if we obtained this public funding we could go to private investors and quickly become international. And that's what happened, we got our sesame, and we didn't expect to go international so quickly. Finally, the support we got at the beginning is the image of the springboard we have, and we can say that it is thanks to this public funding"* **DIR2**.

Innovation consultants bring a perception of the investor that is different from that of other actors. Indeed, the latter characterizes investor capital according to a productive logic. This discernment can highlight problems of alignment of the cognitive schema with regard to the project and its outcome and can lead to conflicts with other stakeholders.

*"In my case, I find it difficult to reason with pure financiers, they have a bit of trouble reasoning other than by money and that's what bothers me. It's very subjective, all these exchanges of information that we can have with them, we are always on : "In reality, their line of conduct is the profitability of the project. In the end, what counts is to work at their own pace in order to reach their objectives, whether they are public or private VCs.*

The relationship between public and private investors is characterized by the common definition of a shared image: innovation. This image constitutes an incentive for the development of formal and informal relations as well as for the exchange of resources. This representation can lead actors with distinct characteristics towards a partnership logic.

*"In our region and our network, innovation is the reason for our joint investments, it is our collective emulsion"* **CIPub 1**.

*"My objective is to continue to develop new companies and new innovations, that's what the investors want, and I also share this point of view*.

However, the complexity of institutional representations in the context of mixed syndication seems to be reduced. As a result, mixed syndication can be a hybrid form of organization that creates value thanks to all the resources and skills exchanged (Murray, 2021).

*"It's obvious that we have differences on the processes and each one has its own way of doing things, but at the end of the day we do the same job and when we work together, we try to do it in the same way. If we have quality investors behind us when it comes to going back into the pot, we don't see the difference at all, so we'll be more inclined to go back to the same people. Look (…) we work almost systematically with them, so we have agreed on a strategic partnership, they are part of the company. That's our advantage, they have skills, we have skills and when we put all that together for a co-investment, we get what we want"* **CIPri 5.**

### 4.3. A "Silicon Valley" effect: between the public actor's quest for legitimacy and productive collaboration.

In this study, we have mainly observed the interactions between public and private VCs with the perspective of mixed syndication relations. However, our analysis seems to indicate a specific characteristic of the public actor which occurs in a context of formal or informal relations when the latter intervenes alone or in syndication. This specificity occupies an essential dimension within complex innovation networks. Indeed, the public VC can express a quest for legitimacy through control. Within a public-private partnership this action could reflect a lack of competence on the part of the public investor who then focuses on this approach (Beuve & Saussier, 2019). In a syndication context, this condition may also represent a means of maintaining one's professional legitimacy that is in line with certain rules (Charreire-Petit & Dubocage, 2018).

*"A manager contacted us to follow up on WCR problems, you immediately said it's ok we refinance (talking to CIPri 4), we preferred to control the whole situation and we presented*

*our partner here with an agreement to commit to a second round, I think you had actually appreciated the study we had done at that time (talking to CIPri 4)"* **CIPub 2.**

However, within an innovation network, this characteristic could also be explained by the significant support of the public actor to companies in the start-up phase. This legitimacy presented to other actors would then have the value of a signal and reputation effect towards the business leaders and also the other actors of this network (Guerini & Quas, 2016. Thus, companies funded by public VCs could benefit from a certification effect with private investors (Guerini & Quas, 2016).

*"We work with them because they are a public partner who facilitates the outcome on certain points, for example the BPI has a preponderant role in the financing of start-ups and SMEs and it is true that we have done quite a few deals with the BPI and yes, it is true that they are recognized for presenting clean and coherent dossiers in terms of expertise when we have to co-invest with them"* **CIPri 7.**

This characteristic implies another phenomenon, that of a "*Silicon Valley*" effect where the complex nature of the networks demonstrates the importance of each agent to the point of making them indispensable (Ferrary & Granovetter, 2009; 2017). This objective can, moreover, be displayed by some VCs and can testify to a possible rooting of VCs within these networks (Granovetter, 1985).

*"In the region, there are not many investors, and knowing that they can refuse to finance a project gives them a little power. It's a good thing, they need us just as much as we need them.*

However, this innovation network dynamic favors productive collaboration between VCs. This may be one of the factors behind the value creation that appears to be achieved through the exchange of financial and/or cognitive resources (Van Aswegen and Retief, 2020).

*"Without cash and without investors, we can do nothing, without precise orientations on the future of the company, innovation, development and the progression of the points of the R&D, we withdraw all perspectives or hope to earn money, we are stronger with two than alone, we will thus collaborate because to innovate is expensive and it is a lot of risks and if we collaborate with our network, we hesitate less "* **CIPri 4.**

In this section, we highlight the specific character of the relationship between VCs, which can be subject to a "*Silicon Valley*" effect. Indeed, within complex innovation networks, VCs try to maintain professional legitimacy with other actors. In this way, the relationship between public and private VCs can be part of a productive collaboration dimension (Stévenot, 2006) where the absence of one of them can represent a difficulty for the whole network (Ferrary & Granovetter, 2017).

## Discussions and conclusions

Our study is one of the first to attempt to understand and define the issues at stake in the relationship between public and private VCs in France. To do so, we have carried out an empirical analysis focused on the interactions between heterogeneous VCs outside and within the context of syndication. In this perspective, we chose a qualitative method of an exploratory nature. We conducted 27 semi-structured interviews with actors of innovation networks identified as: public VCs, private VCs, company managers, innovation advisors and R&D managers. An approach developed by Farrugia (2014), allows us to characterize the relationship between public and private VCs as an "economico-cognitive" approach to networking and innovation (Farrugia, 2014). Indeed, this condition specifies the stakes of the relationship and demonstrates the importance of the capitalization of cognitive and financial resources that allow investors to guarantee ambitions in terms of value creation. Moreover, if some VCs retain observable characteristics in their interventions with companies and managers, the syndication context reveals a hybridization of public and private actors that attenuates these disparities. For all that, our results reveal a process of quest for legitimacy on the part of the public actor characterized by its control role within the public-private partnership (Beuve & Saussier, 2019). If innovation can constitute a shared representation and an incentive as to the intervention of VCs, the network dimension constitutes an essential dimension to the establishment of relations between public and private VCs. These results seem to corroborate the findings made in the research done by Sorenson & Stuart (2001; Pierrakis & Saridakis (2019 page 868) which show that "the structure of social and professional relationships is likely to influence venture capital actors who become aware of promising and early investment opportunities, and timely information about high-quality investment opportunities often reaches a venture capitalist through his or her network". Thus, the dimension of complex networks seems to be decisive in characterizing the relationships between public and private VCs. These spaces or ecosystems are the site of a shared representation: innovation. This constitutes an incentive to pool financial and cognitive resources, particularly in the context of mixed syndication. However, if the hybrid nature of mixed syndication seems to demonstrate that the characteristic features of the different investor-capitalists can be attenuated, certain specificities can persist and disturb the pattern of these relations. Consequently, while public actors seem to possess the competences to encourage the development of the venture capital industry and facilitate "knowledge transfer, tacit knowledge, learning, networking and associated spillover effects" (Pierrakis & Saridakis, 2019, page 868), they are likely to be in search of legitimacy through control . This would be to mask the lack of competence of the public investor who then focuses on this process (Beuve & Saussier, 2019). Thus, in the context of syndication, this condition would allow the public VC to maintain its professional legitimacy which is embedded in the respect of certain rules (Charreire-Petit & Dubocage, 2018).

*Research contributions*

Firstly, this research presents two contributions which concern, on the one hand, the proposed definition of the relationship between public and private VCs and, on the other hand, an update of the concept of *crowding-out* effect. Secondly, the mobilization of the public-private partnership (PPP) approach to try to characterize the relationship between public and private VCs can constitute a contribution in terms of corporate governance. This approach offers a specific insight in order to identify the incentive and/or "disciplining" mechanisms that can be put in place in a context where "*innovation is by essence associated with unpredictable contingencies*" (Hege, 2001, page 9). On another note, this perspective makes it possible to provide certain clarifications for public and private managers, particularly in the case of co-investment or mixed syndication financing. Indeed, if this hybrid form favors the provision of financial and cognitive resources that create value, the question of the distribution of wealth can be a fundamental issue. Thus, the hypothesis of a quest for legitimacy on the part of the public actor could be indicative of a lack of competence on the part of the latter, which would favor the conduct of syndication by a private VC who would be remunerated accordingly. Finally, if this study can highlight a "Silicon Valley" effect that can be characterized by the dynamic and complex nature of a network and the importance of the presence of different heterogeneous actors that compose it, this finding can reveal to public authorities the need to question the innovation policy that can sometimes consist of interventions without considering the location or the presence of other protagonists and lead to bad results. In this way, the public VC could demonstrate its ability "to create a dynamic that adds value to the regional innovation ecosystem by facilitating the production, dissemination and absorption of new knowledge" (Pierrakis & Saridakis, 2019, page 869).

*Boundaries and openings*

First, the limitations of this study concern the qualitative methodology used. The latter needs to be supplemented by a quantitative methodology to verify the effects of the financial intervention of public and private VCs on innovation. Moreover, the fact that we cannot mobilize data from legal actors may constitute a weakness in our study. On the other hand, if our approach focuses only on visible innovation (Djellal & Gallouj, 2018), further research could be conducted regarding the consequences of public and private VCs' intervention on non-visible innovation (Appendix 1) (Djellal & Gallouj, 2018). Finally, the analysis of the relationship between public and private VCs could be the subject of a recent approach developed in the work Frydlinger et al. (2019, page 119), where "the establishment of a formal relational contract could specify mutual goals and establish governance structures to keep the parties' expectations and interests aligned over the long term". To conclude, further research regarding the study of the relationship between public and private equity investors during the Covid 19 crisis could be conducted.

# Annexes

# Annex 1

Figure 1 Representation of the innovation gap, performance gap and policy gap :

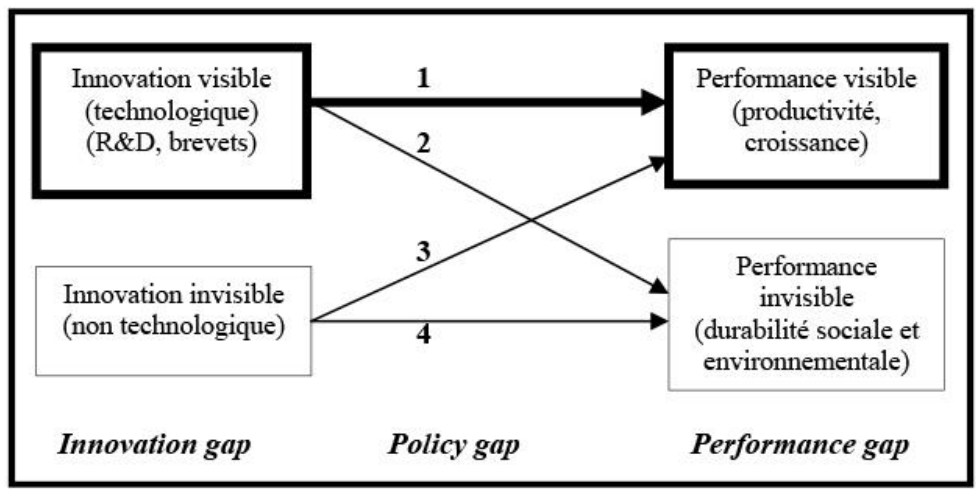

Source: (Djellal and Gallouj, 2010), "Innovation gap, performance gap and policy gap in the service economies" in F. Gallouj, F. Djellal (eds.) (2010). *The Handbook of Innovation and Services: A multidisciplinary perspective* (pp. 653-675), Cheltenham, Edward Elgar.

# Annex 2

Figure 2: Hypotheses explaining the relationship between PPP and innovation :

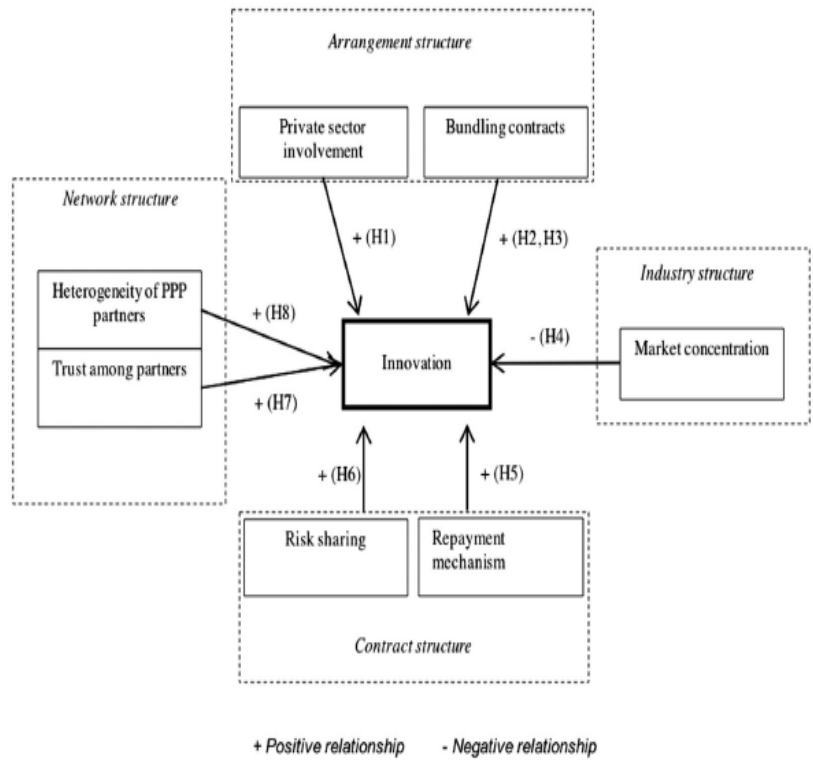

Figure 1. The theoretical framework: relationships between the PPP features and innovation.

Source: Carbonara & Pellegrino (2020), "The role of public private partnerships in fostering innovation" *Construction Management and Economics*, *38*(2), 140-156.

# Annex 3

Figure 3: Representation of the Networking Framework between Venture Capital Funds and the Regional Innovation Ecosystem

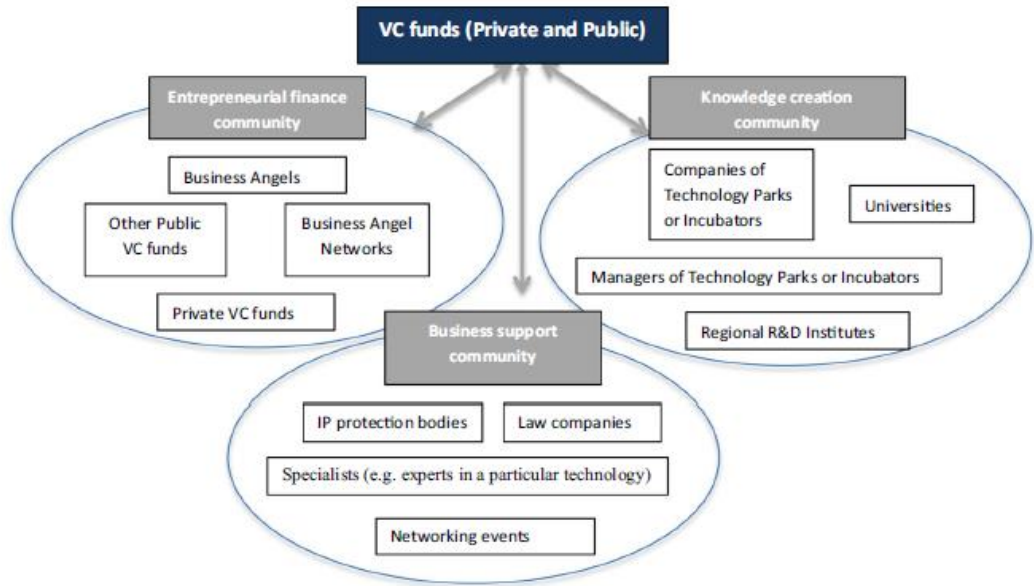

Source: Pierrakis, Y., & Saridakis, G. (2019). The role of venture capitalists in the regional innovation ecosystem: A comparison of networking patterns between private and publicly backed venture capital funds. *The Journal of Technology Transfer*, *44*(3), 850-873.

# Annex 4

Figure 4: Representation of the complex network of the innovative cluster of the silicon valley:

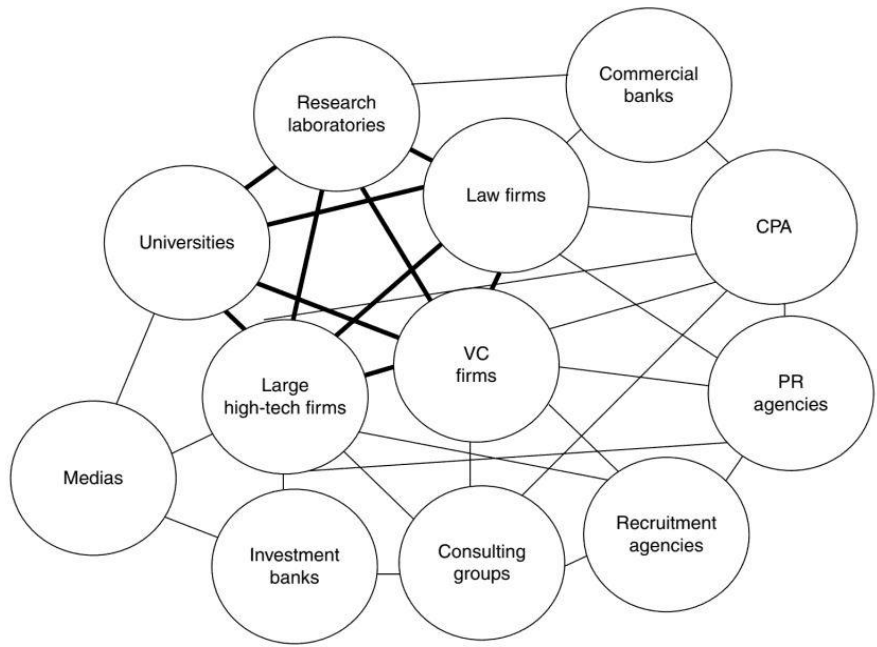

Figure 20.1  Complex networks of Silicon Valley innovative cluster

Source: Ferrary, M., & Granovetter, M. (2017). Social networks and innovation. In *The Elgar Companion to Innovation and Knowledge Creation* (p. 850). Edward Elgar Publishing.